# Enhanced Inference for Finite Population Sampling-Based Prevalence Estimation with Misclassification Errors

Lin Ge, Yuzi Zhang, Lance A. Waller, Robert H. Lyles

Department of Biostatistics and Bioinformatics, Rollins School of Public Health, Emory University, Atlanta, USA

Correspondence email: lge_biostat@outlook.com

# Enhanced Inference for Finite Population Sampling-Based Prevalence Estimation with Misclassification Errors

Epidemiologic screening programs often make use of tests with small, but non-zero probabilities of misdiagnosis. In this article, we assume the target population is finite with a fixed number of true cases, and that we apply an imperfect test with known sensitivity and specificity to a sample of individuals from the population. In this setting, we propose an enhanced inferential approach for use in conjunction with sampling-based bias-corrected prevalence estimation. While ignoring the finite nature of the population can yield markedly conservative estimates, direct application of a standard finite population correction (FPC) conversely leads to underestimation of variance. We uncover a way to leverage the typical FPC indirectly toward valid statistical inference. In particular, we derive a readily estimable extra variance component induced by misclassification in this specific but arguably common diagnostic testing scenario. Our approach yields a standard error estimate that properly captures the sampling variability of the usual bias-corrected maximum likelihood estimator of disease prevalence. Finally, we develop an adapted Bayesian credible interval for the true prevalence that offers improved frequentist properties (i.e., coverage and width) relative to a Wald-type confidence interval. We report the simulation results to demonstrate the enhanced performance of the proposed inferential methods.

1. Introduction

Prevalence estimation is a key component of epidemiological surveillance, allowing healthcare researchers or administrators to make better health policy decisions and take proper actions for disease control and prevention (Rogan & Gladen, 1978). Generally, prevalence monitoring relies on an established diagnostic test to indicate the presence of the disease in individuals. One straightforward way to facilitate valid estimation and inference concerning disease prevalence is to implement a principled sampling design as part of an epidemiologically sound screening program. Several studies (Abdelbasit & Plackett, 1983; BHATTACHARYYA et al., 1979; Birnbaum, 1961; Rao & Scott, 1992) have introduced approaches for point estimation and confidence interval for different types of sampling design. A common issue for sustainable screening programs, however, is that diagnostic results typically rely on an imperfect test or device and are hence subject to misclassification errors. Such misclassifications can be either false positives (e.g., a positive test in a non-diseased individual) or false negatives (e.g., a negative test observed in a diseased patient). As is well known, a proper analytical approach accounting for this misclassification becomes essential for valid prevalence estimation (Bayer et al., 2023; Hallstrom & Trobaugh, 1985).

Two key terms related to the performance of a diagnostic test are *sensitivity* and *specificity*. These are defined as conditional probabilities of test results given the true state (diseased/non-diseased) of the tested individual at the time of the test. To review, the sensitivity of a test denotes the probability of a positive test result given the tested individual is diseased (i.e., Pr[Test positive | diseased]). The specificity of a test denotes the probability of a negative test given the tested individual is non-diseased (i.e., Pr[Test negative | non-diseased]). The probability of a false positive result is Pr[Test positive | non-diseased] = 1 – specificity, while the probability of a false negative result is Pr[Test negative | diseased] = 1 – sensitivity. Note that false positives and false negatives condition on different true states of disease in the tested individual, as do sensitivity and specificity.

Numerous researchers have investigated the implications of applying imperfect diagnostic tests for the purpose of disease monitoring, with a classic early contribution (Bross, 1954) shedding light on the misclassification issue in the context of two-by-two tables and introducing a parameterization to untangle bias in the estimation of crude exposure-disease associations. Other articles (Gastwirth, 1987; Levy & Kass, 1970; Rogan & Gladen, 1978) focus on establishing a bias-corrected estimator of overall prevalence, enabled through the incorporation of known or estimable sensitivity ($Se$) and specificity ($Sp$) parameters characterizing the accuracy of the diagnostic test. To avoid the possibility that the adjusted prevalence estimate could fall outside the (0, 1) range, and particularly to preclude negative estimates when the true prevalence is very small, several researchers (Gaba & Winkler, 1992; Lew & Levy, 1989; Stroud, 1994; van Hasselt et al., 2022; Viana et al., 1993) gravitated to Bayesian approaches wherein prior distributions can define allowable limits for posterior estimates.

The prevalence estimation problem considered in this article is motivated by a disease surveillance system implemented via random sampling of individuals to be tested from a finite target population, such as a clinic patient registry or in-person workforce. We assume use of an imperfect diagnostic test device for which the $Se$ and $Sp$ are known (e.g., specified by the test manufacturer) (Bross, 1954; Levy & Kass, 1970; Sempos & Tian, 2021). The typical bias-corrected prevalence estimator utilizing these known misclassification parameters is readily available (Levy & Kass, 1970). However, to our knowledge no prior study has addressed inference for the true prevalence in the context of screening based on imperfect diagnostic tests when sampling from a finite population.

Here, we propose a valid and enhanced inferential procedure to accompany the common bias-corrected disease prevalence estimate in the finite population scenario. Under the natural assumption that there is a fixed but unknown total number of truly diseased individuals in the target population, the usual finite population correction (FPC) (e.g., Cochran, 1977) applies directly when using a perfect (or "gold-standard") diagnostic test. However, the usual FPC does not apply directly when tests are subject to misclassification, so new inferential methodology is needed for standard error and interval estimation in this case.

A commonly used variance estimator to accompany the standard bias-corrected maximum likelihood (ML) estimate of the true prevalence is well known (Rogan & Gladen, 1978). In what follows, we show that direct application of a standard FPC adjustment (Cochran, 1977) is insufficient for variance adjustment under finite population sampling with a fixed number of true cases. Instead, a subtle but critical extra component of uncertainty exists due to the use of imperfect diagnostic testing and must be accounted for to ensure valid inference. We derive a novel variance estimator incorporating this extra component of variation, enabling the use of a simple Wald-type confidence interval (CI) for the true prevalence. Additionally, we propose an adapted Bayesian credible interval as a way to target more favorable frequentist coverage properties than those obtained by the Wald-type CI in the finite population setting. Finally, we summarize simulation studies quantifying the benefits in performance of the new variance and interval estimation procedures.

## 2. Methods

### *2.1. Preliminary*

We start with a brief review of standard prevalence estimation and inference based on a random sample which applies a "gold-standard" test to each selected individual (i.e., a test with perfect sensitivity and specificity). Suppose we have a random sample of size $n$ from a target population of size $N$. Among those sampled, we record $n^+$ diseased individuals. The simple and familiar prevalence and variance estimators are:

$$\hat{\pi} = n^+/n, \qquad \hat{V}_1(\hat{\pi}) = \frac{\hat{\pi}(1-\hat{\pi})}{n} \qquad (1)$$

Assuming the population is finite with known $N$, the estimated variance of $\hat{\pi}$ incorporates a finite population correction (FPC). We use the version, $FPC = \frac{n(N-n)}{N(n-1)}$, given by Cochran (Cochran, 1977) in pursuit of an unbiased variance estimator, i.e.,

$$\hat{V}_2(\hat{\pi}) = \left[\frac{n(N-n)}{N(n-1)}\right] \frac{\hat{\pi}(1-\hat{\pi})}{n} \qquad (2)$$

A simple Wald-type CI to accompany $\hat{\pi}$ is then immediate, i.e.,

$$\hat{\pi} \pm z_{1-\frac{\alpha}{2}} \sqrt{\hat{V}_2(\hat{\pi})} \qquad (3)$$

Empirical studies indicate that such Wald-type CIs often yield poor coverage properties when the sample size is fixed and small (Agresti & Coull, 1998; Blyth & Still, 1983; Brown et al., 2001; Ghosh, 1979). Recently, Lyles (Lyles et al., 2022) proposed an adjusted Bayesian credible interval approach based on the conjugate beta posterior distribution of the prevalence estimate based on setting a Jeffreys' $Beta(0.5, 0.5)$ prior, and demonstrated better coverage properties for the prevalence estimates when accounting for the associated FPC effect. This new Bayesian credible interval is defined as follows:

$$\left[aQ_{\frac{\alpha}{2}} + b, aQ_{1-\frac{\alpha}{2}} + b\right] \quad, \tag{4}$$

where $a = \sqrt{FPC}$, $b = \hat{\pi}(1-a)$ and $Q_i$ is the 100×$i$-th percentile of the posterior distribution $Beta(n^+ + 0.5, n - n^+ + 0.5)$.

### *2.2. Screening Tests with Misclassification Errors*

Now we consider the scenario in which screening is undertaken by means of an imperfect diagnostic test applied to a random sample of size $n$. We assume the testing procedure is characterized by known manufacturer-specified $Se$ and $Sp$ parameters. In this setting, we define two types of prevalence that can be estimated via random sampling: the test positive frequency ($\pi$) and the true disease prevalence ($\pi_c$), where the latter is of primary interest. Correspondingly, we refer to $N_{c*}$ as the total number of "cases" that would be observed in the event that all individuals were assessed by the imperfect test (including false positive "cases"), and we refer to $N_c$ as the total number of true cases in the population of size $N$. Hence, we have a fixed (unknown) number of true cases ($N_c$) among the population, but a random (unknown) number of total "cases" ($N_{c*}$).

Several authors (Gastwirth, 1987; Levy & Kass, 1970; Rogan & Gladen, 1978) have discussed the bias-corrected true prevalence estimator $\hat{\pi}_c$, i.e.,

$$\hat{\pi}_c = \frac{\hat{\pi} + Sp - 1}{Se + Sp - 1} \quad. \tag{5}$$

The variance estimator of $\hat{\pi}_c$ obtains immediately, as follows:

$$\widehat{Var}(\hat{\pi}_c) = \left(\frac{1}{Se + Sp - 1}\right)^2 \widehat{Var}(\hat{\pi}) \tag{6}$$

In the finite population setting, it may appear reasonable on the surface to use the estimated variance of $\hat{\pi}$ in (2) in conjunction with (6). However, this underestimates the true variance due to the extra variability caused by the fact that it is $N_c$, rather than $N_{c*}$, that is conceptually fixed upon repeated sampling. For a principled derivation, we apply the *Law of Total Variance* to decompose the total variance of $\hat{\pi}$, i.e.,

$$Var(\hat{\pi}) = E[Var(\hat{\pi}|N_{c*})] + Var[E(\hat{\pi}|N_{c*})] \tag{7}$$

Given that the population is closed and finite, the standard FPC is justified for estimating the conditional variance of $\hat{\pi}$ given $N_{c*}$; hence, we replace the first term in (7) by the expression in equation (2). Then for the second term in (7), we apply the variance decomposition rule a second time to characterize the total variance of $E(\hat{\pi}|N_{c*}) = N_{c*}/N$, i.e.,

$$Var[E(\hat{\pi}|N_{c*})] = \frac{Var(N_{c*})}{N^2} = \frac{1}{N^2}\{E[Var(N_{c*}|N_c)] + Var[E(N_{c*}|N_c)]\} \qquad (8)$$

Here $N_{c*}$ is a random variable distributed as a sum of binomials given the total number of true cases $N_c$, governed by the known sensitivity and specificity parameters, i.e.,

$$N_{c*} \sim Bin(N_c, Se) + Bin(N - N_c, 1 - Sp)$$

Therefore, we readily calculate the conditional mean and variance of $N_{c*}$ as follows:

$$E(N_{c*}|N_c) = N_c Se + (N - N_c)(1 - Sp) \qquad (9)$$

$$Var[N_{c*}|N_c] = N_c Se(1 - Se) + (N - N_c)Sp(1 - Sp) \qquad (10)$$

Because $N_c$ is a fixed constant, the variance of (9) is identically zero and the expectation of (10) is itself. Hence, (8) simplifies to the following form,

$$Var[E(\hat{\pi}|N_{c*})] = \frac{1}{N^2}[N_c Se(1 - Se) + (N - N_c)Sp(1 - Sp)]$$
$$= \frac{1}{N}[\pi_c Se(1 - Se) + (1 - \pi_c)Sp(1 - Sp)]. \qquad (11)$$

The estimate of (11) is attained by replacing $\pi_c$ by $\hat{\pi}_c$ in (5). We therefore estimate the total variance of $\hat{\pi}$ in (7) by the summation of (2) and (11), i.e.,

$$\hat{V}_3(\hat{\pi}) = \left[\frac{n(N-n)}{N(n-1)}\right]\frac{\hat{\pi}(1-\hat{\pi})}{n} + \frac{1}{N}[\hat{\pi}_c Se(1-Se) + (1-\hat{\pi}_c)Sp(1-Sp)]. \qquad (12)$$

Finally, the variance estimator associated with the bias-corrected true prevalence estimator $\hat{\pi}_c$ is attained by inserting (12) into (6). Equation (12) is intuitively appealing, as the first term on the right side is the well-known variance that would apply when using a perfect diagnostic test. The second term adjusts that variance upward if the test is imperfect.

When using the bias-corrected prevalence estimator $\hat{\pi}_c$ in (5), one cautionary note relates to the thresholding of $\hat{\pi}_c$. The natural constraints on the problem are that $1 - Sp \leq \hat{\pi} \leq Se$. That is, when a reasonable imperfect diagnostic method is used for screening tests, it should be safe to assume that these constraints apply. Taken together, we can explicitly define a thresholded estimator as the following:

$$\hat{\pi}_c = \begin{cases} 0, & \hat{\pi} \leq 1 - Sp \\ \hat{\pi}_c, & else \\ 1, & \hat{\pi} \geq Se \end{cases} \quad (13)$$

### 2.3. An Adapted Bayesian Credible Interval Approach for Inference

We propose a beta-distribution based posterior credible interval to potentially improve the frequentist coverage properties for the total disease count estimation, accounting for the non-standard variance inflation due to misclassification effects. Similar to (Lyles et al., 2022), our approach implements a scale and shift adjustment to a typical posterior credible interval for the test positive frequency estimator $\hat{\pi}$ based on a Jeffreys' $Beta(0.5,0.5)$ prior and the corresponding conjugate posterior $Beta(n^+ + 0.5, n - n^+ + 0.5)$ distribution. The traditional $100 \times (1-\alpha)\%$ credible interval is defined using the $100 \times \alpha/2$ th and $100 \times (1-\alpha/2)$ th percentiles of this posterior distribution. However, the initial posterior distribution does not account for both finite population effects and misclassification effects as indicated in (12). Therefore, we define a new scale parameter $a' = \sqrt{\hat{V}_3(\hat{\pi})/\hat{V}_1(\hat{\pi})}$ and a shift parameter $b' = \hat{\pi}(1 - a')$. Then all posterior draws $\hat{\pi}_{(j)}, j = 1,2,…,J$ are scaled and shifted, i.e., $\tilde{\pi}_{(j)} = a'\hat{\pi}_{(j)} + b'$. These measures adjust the posterior distribution to have a mean equal to $\hat{\pi}$ and variance approximating that defined in (12). We then take the resulting $100\times(\alpha/2)$th and $100\times(1-\alpha/2)$th percentiles to be a variance-adjusted version of the Jeffreys' interval for $\hat{\pi}$. Therefore, our proposed adapted Bayesian credible interval for use in conjunction with the bias-corrected prevalence estimator $\hat{\pi}_c$ has an analytical form defined by

$$\left[ \frac{(a'Q_{\frac{\alpha}{2}} + b') + Sp - 1}{Se + Sp - 1}, \frac{(a'Q_{1-\frac{\alpha}{2}} + b') + Sp - 1}{Se + Sp - 1} \right] \cap [0,1] \quad , \quad (14)$$

where $Q_i$ is the $100 \times i$-th percentile of the posterior $Beta(n^+ + 0.5, n - n^+ + 0.5)$.

### 3. Simulation Study

We conducted an initial simulation to demonstrate the performance of the bias-corrected prevalence estimator $\hat{\pi}_c$ together with the proposed variance estimator and inferential procedures, under different parameter settings. We define population sizes of $N = 100, 500$ and $1,000$, and examine a range of true disease prevalences ($\pi_c = 0.1, 0.3, 0.5$), sampling rates ($\varphi = 0.1, 0.3, 0.5$), and the known sensitivity and specificity ($Se = 0.9, 0.8; Sp = 0.95, 0.85$). We conducted 5,000 simulations under each setting. All R programs related to the simulation are available in the GitHub site (https://github.com/lge-biostat/Enhanced-Inference-RS-with-Misclassification).

**Table 1 Simulations Evaluating Bias-corrected Prevalence Estimates with N=100[a]**

| | $Se = 0.9, Sp = 0.95$ | | | | | |
|---|---|---|---|---|---|---|
| | $\varphi = 0.1$ | | $\varphi = 0.3$ | | $\varphi = 0.5$ | |
| $\pi_c$ | Mean (SD) [avg. SE] | CI coverage [avg. width] | Mean (SD) [avg. SE] | CI coverage [avg. width] | Mean (SD) [avg. SE] | CI coverage [avg. width] |
| 0.1 | 0.114 (0.109) [0.113] | 76.7%, **97.6%** [0.442], **[0.763]** | 0.101 (0.064) [0.066] | 93.5%, **97.3%** [0.260], **[0.251]** | 0.101 (0.048) [0.048] | 93.9%, **95.1%** [0.189], **[0.183]** |
| 0.3 | 0.298 (0.160) [0.163] | 85.7%, **98.1%** [0.638], **[0.604]** | 0.301 (0.087) [0.088] | 92.8%, **95.7%** [0.344], **[0.331]** | 0.300 (0.063) [0.061] | 94.2%, **94.7%** [0.241], **[0.235]** |
| 0.5 | 0.497 (0.177) [0.179] | 90.3%, **95.9%** [0.702], **[0.629]** | 0.501 (0.095) [0.095] | 93.7%, **95.3%** [0.373], **[0.357]** | 0.499 (0.067) [0.067] | 94.8%, **95.0%** [0.261], **[0.254]** |
| | $Se = 0.8, Sp = 0.85$ | | | | | |
| 0.1 | 0.137 (0.153) [0.190] | 99.0%, **96.9%** [0.754], **[0.872]** | 0.110 (0.097) [0.111] | 98.7%, **95.4%** [0.435], **[0.340]** | 0.104 (0.078) [0.084] | 97.8%, **95.2%** [0.330], **[0.270]** |
| 0.3 | 0.306 (0.213) [0.224] | 96.6%, **97.9%** [0.880], **[0.761]** | 0.301 (0.122) [0125] | 93.0%, **96.0%** [0.492], **[0.462]** | 0.300 (0.093) [0.093] | 95.7%, **95.4%** [0.364], **[0.363]** |
| 0.5 | 0.503 (0.237) [0.237] | 89.2%, **95.4%** [0.930], **[0.809]** | 0.500 (0.131) [0.131] | 95.0%, **95.1%** [0.515], **[0.492]** | 0.499 (0.096) [0.097] | 94.6%, **95.2%** [0.379], **[0.368]** |

[a] SE based on (6) and $\hat{V}_3(\hat{\pi})$ in (12); Wald-based CIs are evaluated (non-bold) along with the proposed adjusted Bayesian credible interval (**bold**)

**Table 1** summarizes the simulation results with $N = 100$. In general, the mean standard error estimated based on our proposed variance estimator provides an excellent match to the empirical standard derivation (SD) of the bias-corrected estimates in most of settings. That is, the novel variance estimator appropriately accounts for the non-standard FPC effects. In contrast, the true variance would tend to be overestimated if ignoring FPC effects altogether, and underestimated if one applied the standard FPC adjustment by plugging (2) into (6). **Figure 1** (described below) presents a clear data visualization to compare these different approaches to variance estimation. In the meantime, we see from **Table 1** that the proposed adjusted Bayesian credible interval generally provides better coverage properties compared to the Wald-type confidence interval, and the mean width of the proposed interval is typically narrower than that of the Wald-type interval.

In reference to **Table 1**, when the misclassification parameters $Se$ and $Sp$ become smaller so that the number of misclassified diagnostic disease status increases in the sample, our proposed estimator remains reliable. However, the point estimates in small prevalence ($\pi_c = 0.1$) and low sampling rate ($\varphi = 0.1$) cases appear to be slightly biased. This is because under the low prevalence setting, the small random sample fails to capture enough "cases" in some iterations, such that $\hat{\pi} < 1 - Sp$. This results in estimates of $\hat{\pi}_c$ truncated by the natural lower bound of zero for the disease prevalence (13). In this situation, the mean of $\hat{\pi}_c$ is expected to be slightly larger than the true value; the standard error also tends to be somewhat inflated in these extreme cases.

**Table 2 Simulations Evaluating Bias-corrected Prevalence Estimates with N=500[a]**

| | \multicolumn{6}{c}{$Se = 0.9, Sp = 0.95$} | | | | | |
|---|---|---|---|---|---|---|
| | \multicolumn{2}{c}{$\varphi = 0.1$} | \multicolumn{2}{c}{$\varphi = 0.3$} | \multicolumn{2}{c}{$\varphi = 0.5$} |
| $\pi_c$ | Mean (SD) [avg. SE] | CI coverage [avg. width] | Mean (SD) [avg. SE] | CI coverage [avg. width] | Mean (SD) [avg. SE] | CI coverage [avg. width] |
| 0.1 | 0.102 (0.053) [0.055] | 92.5%, **95.6%** [0.215], **[0.207]** | 0.100 (0.030) [0.030] | 94.6%, **94.6%** [0.117], **[0.116]** | 0.100 (0.022) [0.022] | 94.8%, **94.8%** [0.085], **[0.084]** |
| 0.3 | 0.301 (0.073) [0.074] | 94.8%, **95.1%** [0.289], **[0.282]** | 0.300 (0.039) [0.039] | 95.2%, **94.7%** [0.154], **[0.152]** | 0.300 (0.027) [0.027] | 95.1%, **95.0%** [0.108], **[0.107]** |
| 0.5 | 0.501 (0.080) [0.080] | 94.3%, **95.0%** [0.314], **[0.305]** | 0.501 (0.042) [0.042] | 94.9%, **95.1%** [0.166], **[0.164]** | 0.500 (0.030) [0.030] | 95.4%, **94.8%** [0.116], **[0.115]** |
| | \multicolumn{6}{c}{$Se = 0.8, Sp = 0.85$} | | | | | |

|   | 0.105 | 98.8%, **94.7%** | 0.101 | 94.5%, **94.7%** | 0.100 | 95.3%, **95.3%** |
|---|---|---|---|---|---|---|
| 0.1 | (0.080) | [0.345], | (0.050) | [0.195], | (0.037) | [0.148], |
|   | [0.088] | **[0.281]** | [0.050] | **[0.181]** | [0.038] | **[0.143]** |
|   | 0.300 | 94.1%, **94.1%** | 0.301 | 94.4%, **94.7%** | 0.300 | 94.4%, **94.7%** |
| 0.3 | (0.104) | [0.397], | (0.057) | [0.220], | (0.041) | [0.163], |
|   | [0.101] | **[0.383]** | [0.056] | **[0.217]** | [0.041] | **[0.161]** |
|   | 0.502 | 94.3%, **95.0%** | 0.499 | 94.4%, **94.8%** | 0.500 | 95.0%, **95.2%** |
| 0.5 | (0.106) | [0.417], | (0.059) | [0.230], | (0.043) | [0.169], |
|   | [0.106] | **[0.405]** | [0.059] | **[0.227]** | [0.043] | **[0.168]** |

a  SE based on (6) and $\hat{V}_3(\hat{\pi})$ in (12); Wald-based CIs are evaluated (non-bold) along with a proposed adjusted Bayesian credible interval (**bold**)

**Table 3 Simulations Evaluating Bias-corrected Prevalence Estimates with N=1,000[a]**

| | $Se = 0.9, Sp = 0.95$ | | | | | |
|---|---|---|---|---|---|---|
| | $\varphi = 0.1$ | | $\varphi = 0.3$ | | $\varphi = 0.5$ | |
| $\pi_c$ | Mean (SD) [avg. SE] | CI coverage [avg. width] | Mean (SD) [avg. SE] | CI coverage [avg. width] | Mean (SD) [avg. SE] | CI coverage [avg. width] |
|   | 0.100 | 93,3%, **94.7%** | 0.100 | 95.3%, **94.8%** | 0.100 | 94.7%, **95.2%** |
| 0.1 | (0.039) | [0.152], | (0.021) | [0.083], | (0.015) | [0.060], |
|   | [0.039] | **[0.150]** | [0.021] | **[0.082]** | [0.015] | **[0.060]** |
|   | 0.300 | 94.4%, **95.0%** | 0.299 | 94.3%, **95.2%** | 0.300 | 95.5%, **95.0%** |
| 0.3 | (0.052) | [0.205], | (0.028) | [0.109], | (0.019) | [0.076], |
|   | [0.052] | **[0.201]** | [0.028] | **[0.108]** | [0.019] | **[0.076]** |
|   | 0.501 | 93.4%, **94.6%** | 0.500 | 95.0%, **94.8%** | 0.500 | 94.6%, **94.4%** |
| 0.5 | (0.057) | [0.222], | (0.030) | [0.118], | (0.021) | [0.082], |
|   | [0.057] | **[0.218]** | [0.030] | **[0.117]** | [0.021] | **[0.082]** |
| | $Se = 0.8, Sp = 0.85$ | | | | | |
|   | 0.100 | 97.4%, **94.6%** | 0.101 | 94.9%, **94.5%** | 0.099 | 95.1%, **95.2%** |
| 0.1 | (0.061) | [0.244], | (0.035) | [0.138], | (0.026) | [0.104], |
|   | [0.062] | **[0.215]** | [0.035] | **[0.135]** | [0.027] | **[0.103]** |
|   | 0.299 | 95.9%, **94.9%** | 0.300 | 95.6%, **94.9%** | 0.300 | 94.1%, **94.3%** |
| 0.3 | (0.071) | [0.281], | (0.039) | [0.155], | (0.030) | [0.115], |
|   | [0.072] | **[0.276]** | [0.040] | **[0.154]** | [0.029] | **[0.114]** |
|   | 0.501 | 94.4%, **94.7%** | 0.500 | 95.5%, **95.0%** | 0.500 | 95.1%, **95.0%** |
| 0.5 | (0.076) | [0.295], | (0.030) | [0.162], | (0.030) | [0.120], |
|   | [0.075] | **[0.290]** | [0.031] | **[0.161]** | [0.031] | **[0.119]** |

a  SE based on (6) and $\hat{V}_3(\hat{\pi})$ in (12); Wald-based CIs are evaluated (non-bold) along with a proposed adjusted Bayesian credible interval (**bold**)

**Table 2** and **Table 3** summarize simulation study results with $N = 500, 1,000$ and demonstrate similar performance characteristics of our proposed variance estimator, as well as the adapted Bayesian credible interval. As expected, when the population size increases to a larger number, the standard errors under each setting become smaller and the width of the estimated intervals becomes narrower as well.

In a subsequent empirical study, we compared the proposed variance estimator to commonly used variance estimators based on the MLE as documented in (Rogan & Gladen, 1978). Here, we assume a random sample for screening testing of size 100 and the true disease prevalence $\pi_c = 0.2$. The bias-corrected prevalence estimator $\hat{\pi}_c$ and each of the available variance estimators were calculated across a wide range of total population size (from 120 to 2,000, with 20,000 simulation runs for each population size). Results were visualized under two types of diagnostic performance settings, one in which $Se = 0.8, Sp = 0.85$ and the other in which $Se = 0.9, Sp = 0.95$. **Figure 1**(A) illustrates results from the first scenario, and **Figure 1**(B) for the second scenario.

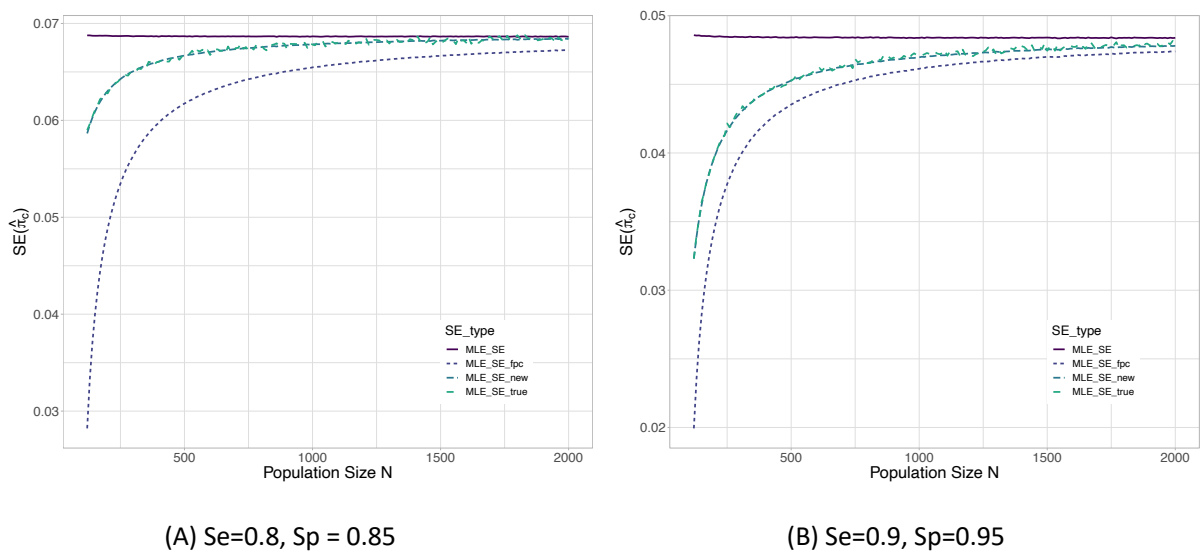

(A) Se=0.8, Sp = 0.85         (B) Se=0.9, Sp=0.95

**Figure 1** Visualization of the Proposed Variance Estimator and Traditional Commonly Used Variance Estimators. In each diagram, from the top to bottom are the averaged standard error calculated via the traditional variance estimator ignoring FPC effects, i.e., inserting $\hat{V}_1(\hat{\pi})$ in (1) into (6) ("MLE_SE", solid line); the averaged standard error calculated via the proposed approach, i.e., inserting $\hat{V}_3(\hat{\pi})$ in (12) into (6) ("MLE_SE_new", blue broken line); the empirical standard derivation of $\hat{\pi}_c$ ("MLE_SE_true", green broken line); the averaged standard error calculated via the traditional variance estimator adjusted by FPC, i.e., inserting $\hat{V}_2(\hat{\pi})$ in (2) into (6) ("MLE_SE_fpc", dotted line).

In each diagram within **Figure 1**, four different lines are plotted. From the top to bottom, the first line indicates the averaged standard error calculated via the traditional MLE-based variance estimator with no FPC adjustment ("MLE_SE"), i.e., inserting $\hat{V}_1(\hat{\pi})$ in (1) into (6). The second line reports the averaged standard error corresponding to our proposed variance estimator, referred to as the new standard error ("MLE_SE_new"), i.e., inserting $\hat{V}_3(\hat{\pi})$ in (12) into (6). The third line shows the empirical standard derivation (SD) of $\hat{\pi}_c$ across simulations ("MLE_SE_true") for each total population size $N$. Finally, the bottom line reports the averaged standard error ("MLE_SE_fpc") based on the traditional variance estimator adjusting for FPC effects, i.e., inserting $\hat{V}_2(\hat{\pi})$ in (2) into (6).

Theoretically and empirically (given that the second and third lines in Figure 1 are nearly indistinguishable), we have shown that the standard error calculated based on our proposed variance estimator provides a close match to the true variability in expectation. **Figure 1** thus provides an intuitive way to illustrate differences between the commonly used variance estimators and our proposed new variance, as well as the similarity between the true sampling SD and our proposed standard error on average. Our proposed estimator ("MLE_SE_new", blue broken line) is bounded above by the traditional variance estimator that ignores any FPC effect, and bounded below by the traditional estimator that applies a standard FPC adjustment as if the number of "cases" ($N_{c*}$), rather than the number of true cases ($N_c$), were fixed. The extra variance due to the misclassification can be substantial, as is the need to properly account for the non-standard FPC effect uncovered in (11), particularly when the finite population is small and the true prevalence also is relatively small. This relates to the fact that (11) is proportional to the inverse of the population size ($\frac{1}{N}$), and hence the four lines in **Figure 1** understandably converge to each other as the population becomes larger.

In **Figure 1** when comparing (A) to (B), we get a sense of the effect caused by changes in the misclassification parameters. As misclassification decreases from left to right, the difference between the "MLE_SE_new" and "MLE_SE_fpc" lines becomes smaller. Ideally, those two lines would be the same if there were no misclassification of diagnosis in the data, because (11) would be zero when the false positive and negative rates are zero. Interestingly, at the same time, we notice that the overestimation of variance associated with ignoring FPC effects altogether is more pronounced in panel (B).

4. **Discussion**

In this article, we propose an enhanced inferential approach in conjunction with bias-corrected prevalence estimation relying on random sampling-based screening tests with misclassification errors in a finite population setting. One timely motivating setting for this line of research is the need to monitor prevalence of infectious diseases such as influenza or COVID-19 among a registered population (e.g., a workforce or retirement community), when the test kits employed are imperfect. We derive a novel corrected version of the variance estimator accompanying the formula (12), which can be used to construct Wald-type confidence intervals. We also propose an adapted Bayesian credible interval incorporating principled shift and scale adjustments, as an alternative to the Wald interval. The essential need for the variance adjustment and inferential procedures provided here stems from the fact that it is the true number of cases, rather than the number of cases identified by the imperfect screening test, that is fixed. As such, there is a non-standard FPC-type effect that must be accounted for. In equation (12), we show that the standard FPC still plays a role, but that there is a critical second term that must be included to properly quantify the variance.

Our simulation results in **Table 1**-**Table 3** suggest that the standard error calculated based on our proposed variance estimator very closely matches with the empirical standard derivation of the estimates across a wide range of settings in this finite population sampling setting. Moreover, we find that the proposed adapted Bayesian credible interval offers a better option for interval estimation, even when the prevalence estimate becomes close to zero, i.e., the estimated positive test frequency $\hat{\pi}$ (the probability of observing a positive test outcome, including both true and false positives) is only slightly smaller than $1 - Sp$. A data visualization summarizing our final set of simulations, provided in **Figure 1**, demonstrates an intuitive explanation for the underestimation variance obtained when applying the standard FPC adjustment when using the bias-corrected MLE in the finite population scenario (i.e., inserting $\hat{V}_2(\hat{\pi})$ in equation (2) into equation (6)). The difference is due to the unavoidable extra variance component due to test imprecision. This extra variance needs to be adjusted for, especially when the population is finite and relatively small.

In this article, we have assumed that the sensitivity and specificity parameters are known and provided by test manufacturers. Nonetheless, there may be potential for misspecification of the sensitivity and specificity parameters due to limited data underlying the manufacturer's estimates, and/or to the divergence between the actual parameters used in practical settings and those determined through laboratory examination. This discrepancy has the potential to introduce bias in disease monitoring among the susceptible population, thus warranting further consideration. In future work, we propose extending the approach outlined here to the setting where we assume $Se$ and $Sp$ are unknown but estimated via external or internal validation data. In practice, this scenario is relevant because these test parameters may be only partially known, particularly when applied to a broader population than used in laboratory validation of the diagnostic tests. Leveraging estimation of these parameters (and associated additional variation) from validation procedures is well understood in the large population setting as demonstrated in (Rogan & Gladen, 1978), but merits further exploration in the smaller, finite population case investigated here.

Finally, as noted by a reviewer, an alternative solution to this problem might be achieved by adopting a fully Bayesian approach (Gaba & Winkler, 1992; Lew & Levy, 1989; Stroud, 1994; van Hasselt et al., 2022; Viana et al., 1993), provided that informative priors on the sensitivity and specificity parameters are available. Such priors would likely need to be based upon validation data, with careful thought given to accommodation of the non-standard finite population correction effects uncovered in the current work.

**Acknowledgement**

This work was supported by the National Institute of Health (NIH)/National Institute of Allergy and Infectious Diseases (P30AI050409; Del Rio PI), the NIH/National Center for Advancing Translational Sciences (UL1TR002378; Taylor PI), the NIH/National Cancer Institute (R01CA234538; Ward/Lash MPIs), and the NIH/National Cancer Institute (R01CA266574; Lyles/Waller MPIs).

**Disclosure statement**

The authors report there are no competing interests to declare.

**Reference**

Abdelbasit, K. M., & Plackett, R. L. (1983). Experimental design for binary data. *Journal of the American Statistical Association*. https://doi.org/10.1080/01621459.1983.10477936

Agresti, A., & Coull, B. A. (1998). Approximate is better than "Exact" for interval estimation of binomial proportions. *American Statistician*. https://doi.org/10.1080/00031305.1998.10480550

Bayer, D. M., Fay, M. P., & Graubard, B. I. (2023). Confidence intervals for prevalence estimates from complex surveys with imperfect assays. *Statistics in Medicine*. https://doi.org/10.1002/sim.9701

BHATTACHARYYA, G. K., KARANDINOS, M. G., & DEFOLIART, G. R. (1979). POINT ESTIMATES AND CONFIDENCE INTERVALS FOR INFECTION RATES USING POOLED ORGANISMS IN EPIDEMIOLOGIC STUDIES1. *American Journal of Epidemiology*, *109*(2), 124–131. https://doi.org/10.1093/oxfordjournals.aje.a112667

Birnbaum, A. (1961). On the Foundations of Statistical Inference: Binary Experiments. *The Annals of Mathematical Statistics*. https://doi.org/10.1214/aoms/1177705050

Blyth, C. R., & Still, H. A. (1983). Binomial confidence intervals. *Journal of the American Statistical Association*. https://doi.org/10.1080/01621459.1983.10477938

Bross, I. (1954). Misclassification in 2 X 2 Tables. *Biometrics*. https://doi.org/10.2307/3001619

Brown, L. D., Cai, T. T., & Das Gupta, A. (2001). Interval estimation for a binomial proportion. *Statistical Science*. https://doi.org/10.1214/ss/1009213286

Cochran, W. G. (1977). *Sampling Techniques, 3rd Edition*. John Wiley.

Gaba, A., & Winkler, R. L. (1992). Implications of Errors in Survey Data: A Bayesian Model. *Management Science*. https://doi.org/10.1287/mnsc.38.7.913

Gastwirth, J. L. (1987). The statistical precision of medical screening procedures: Application to polygraph and AIDS antibodies test data. *Statistical Science*. https://doi.org/10.1214/ss/1177013215

Ghosh, B. K. (1979). A comparison of some approximate confidence intervals for the binomial parameter. *Journal of the American Statistical Association*. https://doi.org/10.1080/01621459.1979.10481051

Hallstrom, A. P., & Trobaugh, G. B. (1985). Specificity, sensitivity, and prevalence in the design of randomized trials:. A univariate analysis. *Controlled Clinical Trials*. https://doi.org/10.1016/0197-2456(85)90118-7

Levy, P. S., & Kass, E. H. (1970). A three-population model for sequential screening for bacteriuria2. *American Journal of Epidemiology*. https://doi.org/10.1093/oxfordjournals.aje.a121122

Lew, R. A., & Levy, P. S. (1989). Estimation of prevalence on the basis of screening tests. *Statistics in Medicine*. https://doi.org/10.1002/sim.4780081006

Lyles, R. H., Zhang, Y., Ge, L., & Waller, L. A. (2022). *A Design and Analytic Strategy for Monitoring Disease Positivity and Case Characteristics in Accessible Closed Populations*. https://doi.org/10.48550/arxiv.2212.04911

Rao, J. N. K., & Scott, A. J. (1992). A Simple Method for the Analysis of Clustered Binary Data. *Biometrics*. https://doi.org/10.2307/2532311

Rogan, W. J., & Gladen, B. (1978). Estimating prevalence from the results of a screening test. *American Journal of Epidemiology*. https://doi.org/10.1093/oxfordjournals.aje.a112510

Sempos, C. T., & Tian, L. (2021). Adjusting Coronavirus Prevalence Estimates for Laboratory Test Kit Error. *American Journal of Epidemiology*. https://doi.org/10.1093/aje/kwaa174

Stroud, T. W. F. (1994). Bayesian analysis of binary survey data. *Canadian Journal of Statistics*. https://doi.org/10.2307/3315826.n2

van Hasselt, M., Bollinger, C. R., & Bray, J. W. (2022). A Bayesian approach to account for misclassification in prevalence and trend estimation. *Journal of Applied Econometrics*. https://doi.org/10.1002/jae.2879

Viana, M. A. G., Ramakrishnan, V., & Levy, P. S. (1993). Bayesian analysis of prevalence from the results of small screening samples. *Communications in Statistics - Theory and Methods*. https://doi.org/10.1080/03610929308831038